\documentclass[referee]{raa}            
\usepackage{amssymb}

\usepackage{graphicx,times}             

\begin{document}

   \title{Spindown of magnetars: Quantum Vacuum Friction?
}

   \volnopage{Vol.0 (200x) No.0, 000--000}      
   \setcounter{page}{1}          

   \author{Xue-Yu Xiong
      \inst{}
   \and Chun-Yuan Gao
      \inst{*}
   \and Ren-Xin Xu
      \inst{}
   }

   \institute{School of Physics and State Key Laboratory of Nuclear Physics and Technology, Peking University,
\\Beijing 100871, China; {\it gaocy@pku.edu.cn}\\
   }

   \date{Received~~2015 April 8; accepted~~2015~~June 20}

\abstract{ Magnetars are proposed to be peculiar neutron stars which could power their X-ray radiation by super-strong magnetic fields as high as $\gtrsim 10^{14}$ G. However, no direct evidence for such strong fields is obtained till now, and the recent discovery of low magnetic field magnetars even indicates that some more efficient radiation mechanism than magnetic dipole radiation should be included.
In this paper, quantum vacuum friction (QVF) is suggested to be a direct consequence of super-strong {\em surface} fields, therefore the magnetar model could then be tested further through the QVF braking.
Pulsars' high surface magnetic field interacting with the quantum vacuum result in a significantly high spindown rate ( $\dot{P}$ ). It is found that QVF dominates the energy loss of pulsars when pulsar's rotation period and its first derivative satisfy the relationship $P^3\cdot \dot{P}>0.63\times10^{-16}\xi^{-4}$s$^2$, where $\xi$ is the ratio of the surface magnetic field over diploe magnetic field.
In the ``QVF $+$ magnetodipole'' joint braking scenario, the spindown behavior of magnetars should be quite different from that in the pure magnetodipole model.
We are expecting these results could be tested by magnetar candidates, especially the low magnetic field ones, in the future.
\keywords{pulsars: general --- radiation: dynamics --- stars: magnetars --- stars: neutron}
}

   \authorrunning{X.-Y. Xiong, C.-Y. Gao \& R.-X. Xu }            
   \titlerunning{Spindown of magnetars: Quantum Vacuum Friction? }   

   \maketitle

%
%
\section{Introduction}           
\label{sect:intro}
Kinematic rotation was generally thought to be the only energy source for pulsar emission soon after the discovery of radio pulsars until the discovery of accretion-powered pulsars in X-ray binaries. However, anomalous X-ray pulsars/soft gamma-ray repeaters (AXPs/SGRs, magnetar candidates) have
long spin periods (thus low spindown power) and no binary companions, which rules out spin and accretion in binary system as the power sources.
The first SGR-giant flare was even observed in 1979 (\cite{Mazets 1979}), and Paczynski (1992) then pointed out that the super-strongmagnetic field may explain the super-Eddington luminosity. AXPs and SGRs are thereafter supposed to be magnetars, peculiar neutron stars with surface/multipole magnetic fields ($10^{14}$ G $\sim 10^{15}$ G) as the energy source, while the initially proposed strong dipole fields could not be necessary (e.g., \cite{Tong 2013}). Moreover, the discovery of low magnetic field magnetars (\cite{Zhou 2014, Rea 2010, Rea 2012, Scholz 2012}) in recent years indicates that some more efficient radiation mechanism than magnetic dipole radiation should be included.
Besides failed predictions and challenges in the magnetar model (\cite{Xu 2007}, \cite{Tong 2011}), one of the key points is: can one obtain direct evidence of the surface strong fields?
Here we are suggesting quantum vacuum friction (QVF) as a direct consequence of the surface fields, and calculating the spindown of magnetar candidates with the inclustion of the QVF effect.

Magnetodipole radiation could dominate the kinematic energy loss
of isolate pulsars (e.g., \cite{Manchester 1977, Dai 1998, Lyubarsky 2001, Morozova 2008}). The derived  braking index $n={\Omega\ddot{\Omega}}/{\dot{\Omega}^{2}}$ ($\Omega$ is the angular velocity of rotation) of a pulsar is expected to be 3
for pure magnetodipole radiation. As a result of observational difficulties, only braking indices $n$ of a few rotation-powered pulsars are obtained with some certainty (\cite{Yue 2007, Livingstone 2007}, http://www.atnf.csiro.au/research/pulsar/psrcat/). They are PSR J1846-0258
($n = 2.65 \pm 0.01$), PSR B1509-58 ($n = 2.839 \pm 0.001$), PSR J1119-6127 ($n = 2.91 \pm 0.05$), PSR B0531+21 (the Crab pulsar, $n = 2.51 \pm 0.01$)), PSR B0540-69 ($n = 2.140 \pm 0.009$) and PSR B0833-45 (the Vela pulsar, $n = 1.4 \pm 0.2$). These observed breaking indices are all remarkably smaller than the value of $n = 3$, which may suggest that other spin-down torques do work besides the energy loss via dipole radiation (\cite{Xu 2001, Beskin 1984, Ahmedov 2012, Menou 2001, Contopoulos 2006, Alpar 2001, Chen 2006, Ruderman 2005, Allen 1997, Lin 2004, Tong 2014, Tong 2015}).

Recently, the research of Davies et al. shows that the
QVF effect could be a basic electromagnetic
phenomenon (\cite{Davies 2005, Lambrecht 1996, Pendry 1997, Feigel
2004, Tiggelen 2006, Manjavacas 2010}). If the quantum vacuum friction exists, the dissipative energy by QVF would certainly
be from rotational kinetic energy of pulsar. The loss
of rotational kinetic energy of pulsar by QVF may also transform into
pulsar's thermal energy or the energy of pulsar's radiating photons
which might not be isotropic. This is the same argument as in the
work of Manjavacas et al. (2010), in which the authors argue that at zero temperature, the friction produced on
rotating neutral particles by interaction with the vacuum
electromagnetic fields transforms mechanical energy into light emission and produces particle heating. Pulsar may transfer its
angular momentum to the vacuum when pulsars rub against quantum
vacuum since the angular momentum is conserved. In this case, vacuum may work as an standard medium (\cite{Dupays 2008}).
Dupays et al. (2008, 2012) even calculated the energy loss due to pulsars'
interaction with the quantum vacuum by taking account of quantum
electrodynamics (QED) effect in high magnetic field. The calculations indicate that when the pulsars' magnetic
field is high, QVF would also play an important role to cause the
rotation energy loss of pulsars. Thus, it is necessary to take QVF
into the rotation energy loss of pulsars, especially for highly
magnetized pulsars on surface, like magnetars.

In this paper we assume that pulsar interacts with quantum vacuum as
in the work of Dupays et al. (2008) and consider the
difference between the surface/toroidal magnetic field and dipole/poloidal magnetic field. The braking
indices for pure QVF radiation and surface magnetic field of magnetars for the ``QVF $+$ magnetodipole'' joint braking model are calculated.

The paper is organized as following. After an introduction, we
deduce the relation between the dipole magnetic field and the braking index of magnetars
in the second section. The calculated results and analysis are
presented in the third section. Finally, conclusions and discussions are presented.


\section{Spindown and braking index of magnetars}
\label{sect:model}
A pulsar has the power of magnetodipole radiation of
\begin{equation}\dot{E}_{\rm{dip}}=-\frac{2}{3}c^{-3}\mu^{2}\Omega^{4},\label{1}\end{equation}
where \begin{equation}\mu=\frac{1}{2} B_{\rm dip}R^3 \sin\theta\label{mu}\end{equation} is magnetic dipolar moment and $c$ is the speed of light in vacuum, $B_{\rm dip}$ is the dipole magnetic field, $R$ is the pulsars' radius, $\theta$ is the inclination angle.
For general pulsars, surface magnetic field approximately equal to dipole magnetic field because multipole magnetic field attenuate to little. However, for magnetars there is a surplus of attenuate multipole magnetic field as its extraordinarily strong surface magnetic field. So magnetars' surface magnetic field $B_{\rm surf}$ include dipole magnetic field
$B_{\rm dip}$ and multipole magnetic field. We suppose that the ratio of surface magnetic field
and dipole magnetic field
\begin{equation}
\xi=\frac{B_{\rm surf}}{B_{\rm dip}}
\end{equation}
is a constant. The pulsar rubs against the quantum vacuum and then loses its
rotation kinetic energy (\cite{Dupays 2008}) of
\begin{equation}\dot{E}_{\rm qvf}\simeq-\alpha\frac{3\pi}{16}
\frac{R^{4}\sin^{2}\theta}{c B_{c}^{2}}\frac{B_{\rm surf}^{4}}{P^{2}},\label{2}
\end{equation}
where $\alpha=e^{2}/\hbar c\simeq1/137$ is the coupling constant of
electromagnetic interaction, $B_{c}=4.4\times10^{13}G$ is the QED
critical field and
$P=2\pi/\Omega$ is the spin period.

The pulsars' typical radius $R=10^6$cm is adopted. Set inclination angle
$\theta=90^{\circ}$ for the sake of simplicity. Considering the relation (\ref{mu}) between
the magnetic moment of pulsars and magnetic field in polar region of
pulsars, we can obtain the ratio
of the energy loss due to QVF over that due to magnetodipole radiation
 \begin{equation}\frac{\dot{E}_{\rm qvf}}{\dot{E}_{\rm dip}}
 =7.69\times10^{-24} B_{\rm dip}^2 P^2{\xi}^4.\label{4}
 \end{equation}

Assuming the pulsars' rotation energy loss coming from both magnetodipole
radiation and QVF, i.e. $\dot{E}=\dot{E}_{\rm dip}+\dot{E}_{\rm qvf}$, the
total energy loss of pulsars are given by
\begin{equation}\dot{E}\simeq-\frac{2\mu^2\Omega^{4}}{3c^{3}}-\alpha\frac{3\pi}{16}\frac{\sin^{2}\theta}{c B_{c}^{2}
}\frac{B_{\rm dip}^{4}R^{4}}{P^{2}}{\xi}^4.
\end{equation}
From the pulsars' rotation energy loss $\dot{E}=I\Omega\dot{\Omega}$, where $I$ is the inertia of momentum with typical value
$I=10^{45}$g$\cdot$cm${^2}$, we can obtain a relationship
between pulsar's period and the period derivative with respect to
time
\begin{equation}\dot{P}=\frac{2{\pi}^2R^6\sin^{2}\theta}{3c^3I}\frac{B_{\rm dip}^2}{P}
+\frac{3\alpha R^4\sin^{2}\theta}{64\pi B_c{}^2Ic}\frac{}{}B_{\rm dip}^4P{\xi}^4.\label{6}
\end{equation}

Using the relation
of $\Omega$ and $P$, the braking index can be obtained
\begin{equation}n=\frac{1}{\dot{P}}\left(\frac{2\pi^2  R^6\sin^{2}\theta}{
Ic^3}\frac{B_{\rm dip}^2}{P}+
\frac{3\alpha}{64\pi}\frac{{R^4\sin}^2\theta}{B_{c}^{2}Ic}B_{\rm dip}^4 P{\xi}^4\right).
\end{equation}
Numerically, the braking index can be written as
\begin{equation}n=\frac{7.31+f(B_{\rm dip},P,\xi)}{2.44+f(B_{\rm dip},P,\xi)},\label{7}
\end{equation}
where
\begin{equation}f(B_{\rm dip},P,\xi)=18.75B_{\rm dip,12}^{2}P^{2}\xi^{4}\label{8}
\end{equation}
with $B_{\rm dip,12}=10^{-12} B_{\rm dip}$. We can also express the ratio
of the energy loss due to QVF over that due to magnetodipole radiation by pulsar's period($P$) and period derivative($\dot{P}$) from equation (\ref{4}) and (\ref{6})
\begin{equation}\frac{\dot{E}_{\rm qvf}}{\dot{E}_{\rm dip}}
 =7.69\times10^{-24} \left(-\frac{8768}{9c^2}\pi^{3}R^2 B_{c}^{2}+\sqrt{(\frac{8768}{9c^2}\pi^{3}R^2 B_{c}^{2})^2+\frac{8768}{3R^4}\pi IcB_{c}^2\xi^4P^3\dot{P}}\right).\label{9}
\end{equation}
Numerically, the above equation can be written as
\begin{equation}\frac{\dot{E}_{\rm qvf}}{\dot{E}_{\rm dip}}
 =-\frac{1}{2}+\sqrt{\frac{1}{4}+3.16\times10^{16}\xi^4P^3\dot{P}}.\label{10}
\end{equation}

\section{The numerical results}
\label{sect:results}

The periods of observed pulsars are distributed mainly in the range
from $0.1$s to $5$s (The ATNF Pulsar Catalogue:
http://www.atnf.csiro.au/research/pulsar/psrcat/). Using Eq. (\ref{4}) we
plot the ratio of $\dot{E}_{\rm qvf}/\dot{E}_{\rm dip}$, as a
function of the period $P$ in Fig. 1 for $\xi=10$ and in Fig. 2 for $\xi=100$.
From Fig.1 we can see that QVF may play an important role when the dipole magnetic field is higher than
$\sim 10^{10}$G for pulsars whose period are between $0.1$s and 1s.
Most of observed pulsar's magnetic field derived from pure
magnetodipole radiation are in the region $10^{11}-10^{13}$G,
however, if QVF is included in pulsars' energy loss, the
derived magnetic field could be lower. Thus it is necessary to independently
measure the magnetic field of pulsars so that we can judge whether QVF has
important contribution to pulsars' rotation energy loss.

From Fig.2 we can see that QVF may play an important role when
pulsars dipole magnetic field $B_{\rm dip}>10^{10}$ for most pulsars' braking.
For millisecond pulsars the derived magnetic field from
magnetodipole radiation is already so low ($B_{\rm dip}<10^{10}$G) that we can
neglect the QVF's contribution to its rotation energy loss, but for
magnetars the derived magnetic field from QVF is already so high
($B_{\rm dip}>10^{12}$G) that we have to consider the QVF's contribution.
We can also express the ratio of the energy loss due to QVF over that due to magnetodipole
radiation by pulsar's period($P$) and period derivative($\dot{P}$) as shown in Eq. (\ref{10}).
From this equation we can obtain that QVF dominates the energy loss of pulsars when pulsar's rotation period and its first derivative satisfy the relationship $P^3\cdot \dot{P}>0.63\times10^{-16}\xi^{-4}$s$^2$, where $\xi$ is the ratio of the surface magnetic field over diploe magnetic field.
According to above relationship and current observed data for confirmed magnetars (see Table 1) QVF will dominate the rotation energy loss in all of the magnetars' spindown.

\begin{figure}
   \centering
   \includegraphics[width=0.9\textwidth, angle=0]{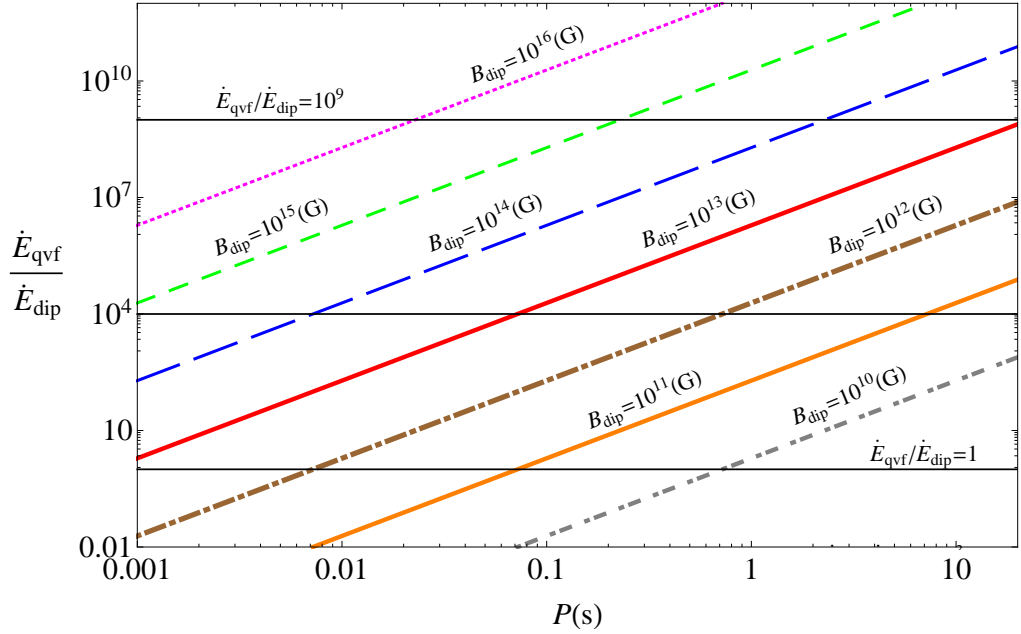}
   \caption{The ratio of a pulsar's energy loss rate from QVF over that from magnetodipole radiation, as a function of period, where $\xi =10$.}
   \label{Fig:demo1}
   \end{figure}

   \begin{figure}
   \centering
   \includegraphics[width=0.9\textwidth, angle=0]{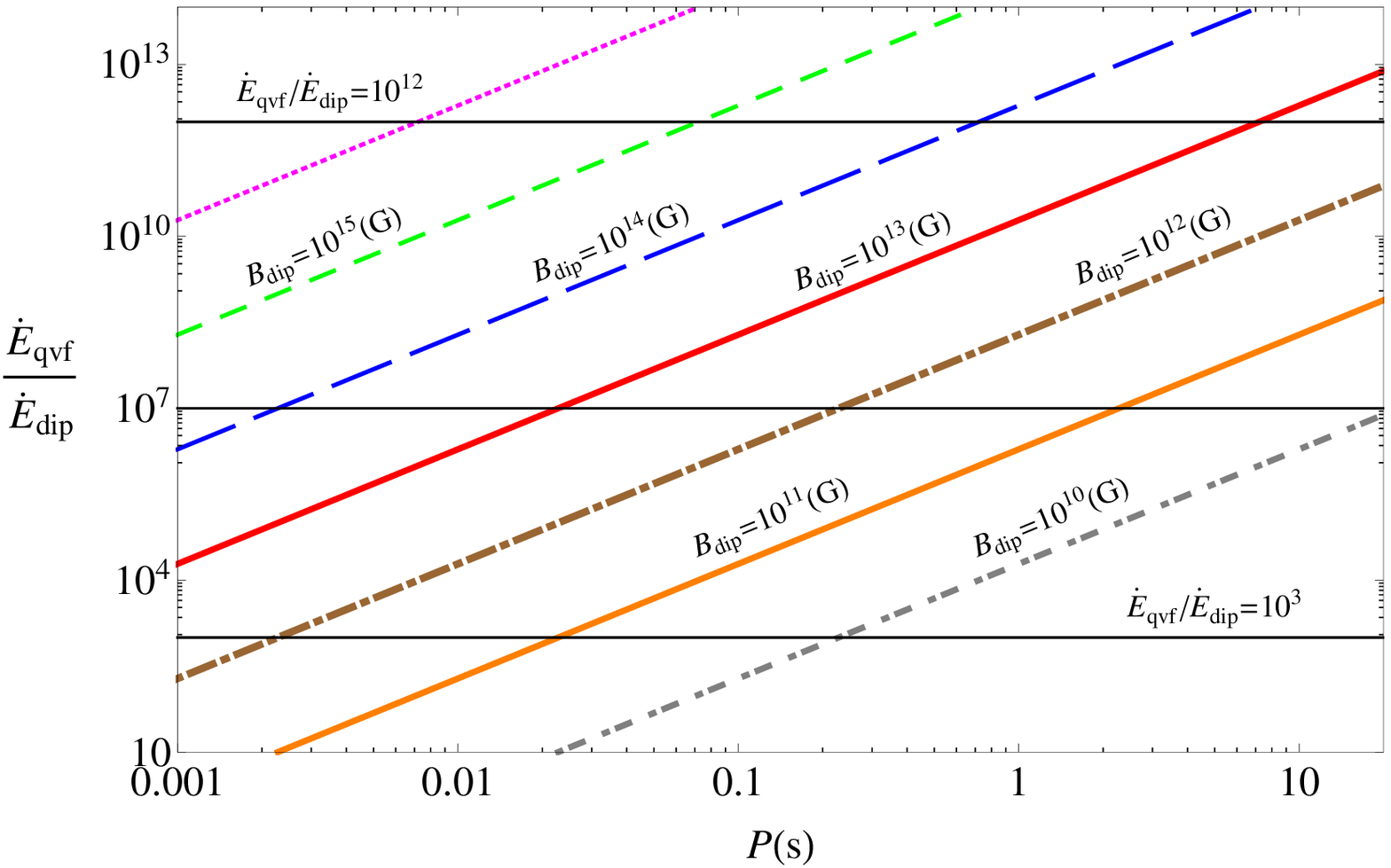}
   \caption{The ratio of a pulsar's energy loss rate from QVF over that from magnetodipole radiation, as a function of period, where $\xi =100$.}
   \label{Fig:demo2}
   \end{figure}

\begin{table}
\begin{center}
\caption[]{The parameters and the inferred magnetic field of
magnetars. The magnetars' data of period ($P$), the period
derivative ($\dot{P}$) and dipole magnetic field ($B_{\rm dip}$) are
from the McGill SGR/AXP Online Catalog
(http://www.physics.mcgill.ca/~pulsar/magnetar/main.html). The last
column of table, $B_{\rm dip}^{\rm inf}$, is inferred magnetic field from our model based on
both magnetodipole radiation and QVF.}\label{Tab:publ-works}


 \begin{tabular}{cccccc}
  \hline\noalign{\smallskip}
Name&$P$(s)&$\dot{P}$($10^{-11}$s/s)&$B_{\rm dip}$($10^{14}$G)&$B_{\rm dip}^{\rm inf}$($10^{11}$G, ~$\xi=10$)&$B_{\rm dip}^{\rm inf}$($10^{10}$G,~$\xi=100$)\\
\hline\noalign{\smallskip}
CXOU J010043.1-721134&8.020392(9)&1.88(8)&3.9&5.946&5.946\\
4U 0142+61&8.68832877(2)&0.20332(7)&1.3&3.342&3.342\\
SGR 0418+5729&9.07838827&$<0.0006$&$<0.075$&$<0.770$&$<0.770$\\
SGR 0501+4516&5.76209653&0.582(3)&1.9&4.817&4.817\\
SGR 0526-66&8.0544(2)&3.8(1)&5.6&7.082&7.082\\
1E 1048.1-5937&6.4578754(25)&~2.25&3.9&6.565&6.565\\
1E 1547.0-5408&2.06983302(4)&2.318(5)&2.2&8.791&8.791\\
PSR J1622-4950&4.3261(1)&1.7(1)&2.7&6.766&6.766\\
SGR 1627-41&2.594578(6)&1.9(4)&2.2&7.905&7.905\\
CXO J164710.2-455216&10.6106563(1)&0.083(2)&0.95&2.541&2.541\\
1RXS J170849.0-400910&11.003027(1)&1.91(4)&4.6&5.516&5.516\\
CXOU J171405.7-381031&3.825352(4)&6.40(5)&5.0&9.718&9.718\\
SGR J1745-2900&3.76363824(13)&1.385(15)&2.3&6.655&6.655\\
SGR 1806-20&7.6022(7)&75(4)&24&15.145&15.145\\
XTE J1810-197&5.5403537(2)&0.777(3)&2.1&5.229&5.229\\
Swift J1822.3-1606&8.43772106(6)&0.00214(21)&0.14&1.078&1.078\\
SGR 1833-0832&7.5654091(8)&0.439(43)&1.8&4.194&4.194\\
Swift J1834.9-0846&2.4823018(1)&0.796(12)&1.4&6.430&6.430\\
1E 1841-045&11.7828977(10)&3.93(1)&6.9&6.494&6.494\\
3XMM J185246.6+003317&11.55871346(6)& $<0.014$& $<0.41$&$<1.594$&$<1.594$\\
SGR 1900+14&5.19987(7)&9.2(4)&7.0&9.856&9.856\\
1E 2259+586&6.9789484460(39)&0.048430(8)&0.59&2.466&2.466\\
PSR J1846-0258&0.32657128834(4)&0.7107450(2)&0.49&10.379&10.379\\   \\
  \noalign{\smallskip}\hline
\end{tabular}
\end{center}
\end{table}

Substituting the observed value of $\dot{P}$ and $P$ into Eq. (\ref{6}),
the magnetic field of pulsars can be calculated. We compute the
currently confirmed magnetars' magnetic field and list the results
in the last column $B_{\rm dip}^{\rm inf}$ of Table 1. The fourth column
$B_{\rm dip}$ is derived from pure magnetodipole radiation.  The calculated results
manifest that the derived dipole magnetic field $B_{\rm dip}$ from pure
magnetodipole radiation is about $10^3$($\xi=10$) and $10^4$($\xi=100$)
 times larger than $B_{\rm dip}^{\rm inf}$ obtained by combining QVF and
 magnetodipole radiation. And the derived surface magnetic field $B_{\rm surf}$ from pure
magnetodipole radiation is about 100
times larger than $B_{\rm dip}^{\rm inf}$ inferred by combining QVF and
 magnetodipole radiation for both $\xi=10$ and $\xi=100$.

If $\dot{E}=\dot{E}_{\rm QVF}$, from Eq. (\ref{2}) we can obtain $\dot \Omega\propto \Omega^1$, therefore braking index $n=1$ for pulsar's spindown by pure QVF. Eq. (\ref{7}) show that pulsar's braking index is between $1\sim3$ in the `QVF $+$ magnetodipole'' joint braking scenario. Magnetars have strong surface magnetic field, longer rotation period and bigger $\xi$, so magnetars have bigger $f(B_{\rm dip},P,\xi)$ function value (see Eq. (\ref{8})) which result in QVF dominating magnetars' braking and its braking indices being about $3$. However, for some low magnetic field millisecod pulsar, minor $f(B_{\rm dip},P,\xi)$ function value lead to magnetodipole radiation becoming main energy loss way in its spindown and its braking index is about $3$.
Considering pulsar's spindown by both QVF and magnetodipole radiation, we
use Eq. (\ref{7}) to calculate the braking indices of magnetars. The results
show that all the magnetars' braking indices are around 1 for both $\xi=10$ and $\xi=100$. In the future, the model could be tested by comparing the calculated results to observed braking
indices. This comparison can also provide further information to understand QVF.

\section{Conclusions and Discussions}
\label{sect:conclusion}
We investigate pulsar's rotation energy loss from QVF and compare it with that from magnetodipole radiation in the different magnetic
field range and different period range. We find that if the ratio $\xi$ of the surface magnetic field over dipole
 magnetic field is fixed to $10(100)$, QVF could play
a critical role for pulsars' braking when $B_{\rm surf}\cdot P>10^{11}(10^{10})$G$\cdot$s,
while it can be ignored when $B_{\rm surf}\cdot P<10^{10}(10^{9})$G$\cdot$s. Magnetars may have high surface magnetic field and long
period ($B_{\rm surf}\cdot P\gg 10^{12}$G$\cdot$s) if the value of magnetic
field is inferred by pure classical magnetodipole radiation. Therefore it is necessary to
consider magnetars' rotation energy loss by both
magnetodipole radiation and QVF.

We consider the difference between the surface magnetic field and
dipole magnetic field of pulsars and compare the energy loss rate of pulsars
due to magnetodipole
radiation to that due to QVF. The results show that when a pulsar
has a strong magnetic field or a long period
($B_{\rm surf}\cdot P>10^{11}$G$\cdot$s for $\xi =10$,
 $B_{\rm surf}\cdot P>10^{10}$G$\cdot$s for $\xi =100$ ), comparing to QVF, the energy loss by magnetodipole radiation can be ignored, while when pulsars have weak
magnetic field or short period
($B_{\rm surf}\cdot P<10^{10}$G$\cdot$s for $\xi=10$,
$B_{\rm surf}\cdot P<10^{9}$G$\cdot$s for $\xi=10$) the
QVF can be negligible. We consider that rotation energy loss of
magnetars is the sum of the energy loss due to QVF and that due to
magnetodipole radiation. Based on this joint mechanism of energy
loss, the surface magnetic field of magnetars and braking indices
are calculated. Our work indicates that when QVF is included in
the process of rotation energy loss, the surface magnetic field
of magnetars is $10-100$ times lower than that in pure
magnetodipole radiation model. In this joint braking model QVF dominates the energy loss of pulsars when pulsar's rotation period and its first derivative satisfy the relationship $P^3\cdot \dot{P}>0.63\times10^{-16}\xi^{-4}$s$^2$, where $\xi$ is the ratio of the surface magnetic field over diploe magnetic field. Also, we obtain the braking index of
magenetars is around 1 in the joint braking model. The efficiency of rotation energy losses
generated by QVF in magnetars is very high compared to magnetic dipole radiation. Smaller magnetic field can
 generate a greater rotation energy loss by QVF comparing to magnetic dipole
 radiation. This may explain why magnetars which have great X-ray luminosity
  and low magnetic field (\cite{Zhou 2014, Rea 2010, Rea 2012, Scholz 2012}).

We are expecting the results presented could be tested by X-ray observations of magnetar candidates, especially for the low magnetic field ones.
 X-ray data accumulated in space advanced facilities could show both timing and luminosity features for magnetars, and a data-based research would be necessary and interesting.
Summarily, further observations for magnetars in the future would test our joint braking model as well as help us understand QVF in reality.

\begin{acknowledgements}
The authors thank Yue You-ling, Feng
Shu-hua, Liu Xiong-wei and Yu Meng for helpful discussions. This work
is supported by the National Natural Science Foundation of China
(11225314), XTP XDA04060604, and SinoProbe-09-03 (201311194-03).
\end{acknowledgements}


\label{lastpage}

\end{document}